# Optimizing Coordinative Schedules for Tanker Terminals: An Intelligent Large Spatial-Temporal Data-Driven Approach - Part 1

*Deqing Zhai[a,b,\*], Xiuju Fu[a,\*], Xiao Feng Yin[a], Haiyan Xu[a], Wanbing Zhang[a] and Ning Li[a]*

[a]*1 Fusionopolis Way, Institute of High Performance Computing, Agency for Science, Technology and Research, Singapore 138632*

[b]*50 Nanyang Avenue, School of Electrical and Electronic Engineering, Nanyang Technological University, Singapore 639798*

## Abstract

In this study, a novel coordinative scheduling optimization approach is proposed to enhance port efficiency by reducing average wait time and turnaround time. The proposed approach consists of enhanced particle swarm optimization (ePSO) as kernel and augmented firefly algorithm (AFA) as global optimal search. Two paradigm methods of the proposed approach are investigated, which are batch method and rolling horizon method. The experimental results show that both paradigm methods of proposed approach can effectively enhance port efficiency. The average wait time could be significantly reduced by 86.0% - 95.5%, and the average turnaround time could eventually save 38.2% - 42.4% with respect to historical benchmarks. Moreover, the paradigm method of rolling horizon could reduce to 20 mins on running time over 3-month datasets, rather than 4 hrs on batch method at corresponding maximum performance.

[\*]Corresponding author(s):

Email address (es): dzhai001@e.ntu.edu.sg (Deqing Zhai), fuxj@ihpc.a-star.edu.sg (Xiuju Fu)

# Introduction

Maritime study is one of many long history studies. It is a multi-disciplinary academic field that covers almost every profession having to do with the sea [1]. Hence, little enhancement over maritime industry would generate a big leap on worldwide logistic networks, so that it is important to optimize the networks comprehensively by various state-of-the-art studies of machine learning and optimization techniques. Optimal networks are chains that make agents in the networks smoothly transit with less cost required. As a matter of fact, they would eventually reduce various costs, such as operation cost, fuel cost, travel cost, etc.

Adopted by International Maritime Organization (IMO) in 2000, a newly revised requirement for maritime navigational systems shows that a vessel is to carry automatic identification systems (AIS) that can be capable of broadcasting static, dynamic, voyage-cargo related knowledge from vessels to vessels, from vessels to coastal authorities or even to satellite communication links automatically [2]. General knowledge broadcast by AIS transceivers can be categorized in Figure 1.

| Dynamic Voyage-Cargo Knowledge | | | | |
|---|---|---|---|---|
| Item | Last Port | Next Port | ETA | Cargo | Size |
| Type | Character | Character | Numeric (UTC) | Character | Numeric |

| Static Vessel Knowledge | | | |
|---|---|---|---|
| Item | IMO | MMSI | Call Sign | Name |
| Type | Numeric | Numeric | Character | Character |
| Item | Type | Length | Breadth | Location of Antenna Fixed |
| Type | Character | Numeric | Numeric | Numeric |

| Dynamic Vessel Knowledge | | | |
|---|---|---|---|
| Item | Position | Time Stamp | Course over Ground | Speed over Ground |
| Type | Numeric (Lat, Lon) | Numeric (UTC) | Numeric | Numeric |
| Item | Heading | Navigational Status | Rate of Turn | Draught |
| Type | Numeric | Character | Numeric | Numeric |

**Figure 1: AIS General Knowledge**

According to the regulations of IMO as shown in Figure 2, all cargo vessels that are engaged on international voyages with gross tonnage 300 and above are required to fit AIS devices. Moreover, all cargo vessels, which are not engaged on international voyages with gross tonnage 500 and above, are required to equip with AIS devices on board. Lastly, all passenger vessels are required to have AIS devices on board irrespective of vessel sizes [2].

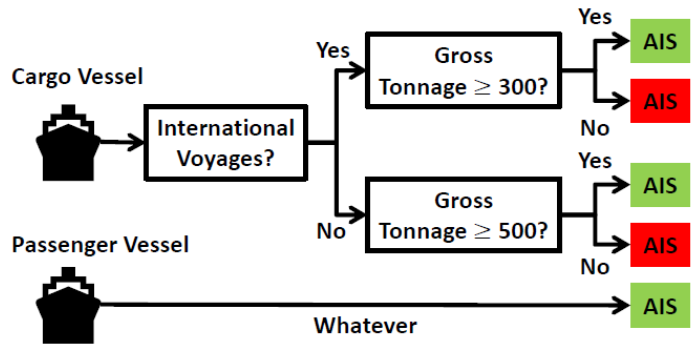

Figure 2: AIS Device Engagements

With the AIS embedded on board, the AIS transceiver transmits data in standard format every 2 - 10 seconds depending on the speed of vessel while under way, and transmits data every 3 minutes while in anchor [3]. There are about more than 50,000 different types of vessels operated all over the world in 2018 [4]. Thus, these generate millions of large spatial-temporal data every few seconds among vessels all over the world. On the basis of big spatial-temporal data, there is a number of useful hidden knowledge that can be figured out through data mining and data analytics. The useful hidden knowledge could be handy databases for relevant researches, such as traffic analysis, anomaly detection, route pattern detection, collision avoidance, route planning, scheduling optimization, etc. However, how do we standardize and store this large spatial-temporal data is still one of many challenging problems, and it is not the main scope of this paper. With obtaining this large spatial-temporal data, this study proposes a systematical optimization approach on coordinative scheduling in tanker terminals as scopes, and the effectiveness of proposed approach has been validated by historical benchmarks with two proposed methods as well.

This paper is organized as follows: Section 2 reviews relevant literature regarding on scheduling optimization in maritime traffic. Section 3 formulates the problems that are to be tackled. The proposed methodology is followed up in Section 4. Section 5 presents the results and corresponding discussion on the basis of proposed methodology. Final conclusion, limitation and future direction are drawn in Section 6.

## Literature Review

Back in 1999, Z. Wang apparently showed that different activity-cargo types from different mixture of vessels in all directions are one of very challenging problems [5]. There are four basic components of cargo operation, which are ship operation, terminal operation, storage operation and receipt/delivery operation as shown in Figure 3.

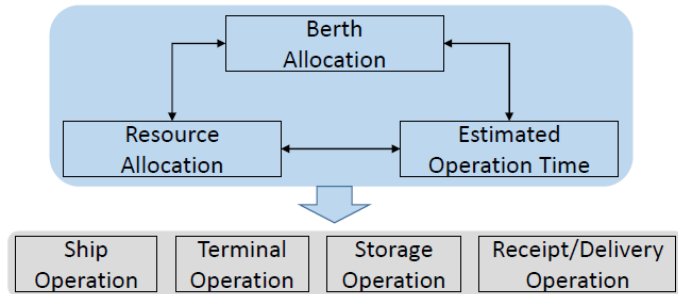

Figure 3: Shipping Scheduling and Planning

In order to make these four operations efficiently work together, a coordinative scheduling is one of necessary tasks [5]. Different from other long-term aspects in maritime, the coordinative scheduling is normally having a short-term time scale, such as only a few days or weeks in advance for smoothly completing cargo operations with minimum in-port time elapsed in the meantime. Since these operations are not mutually exclusive in terms of operation sequences, they impact on each other and affect the respective operations, which would lead to ineffective and changing schedules for overall shipping scheduling and planning. In the study of E. Nishimura et al., a heuristic algorithm (genetic algorithm, GA) was proposed to effectively generate numerical simulation solutions with remarkable reduction of computational efforts for terminal operation section in shipping scheduling and planning [6]. However, the scopes of berths and vessels are quite limited in the study with only no more than 10-berth scenario. Since the computational efforts of GA are exponentially increasing with agent chromosomes, feasible solutions could not be obtainable or they could cost ineffective computational efforts if the scale of problem expands. Zhang et al. proposed a solution by coordinating to schedule channels and berths in port operations. The model focuses on the considerations of scheduling order, traveling direction and distance of a berth. A mathematical objective function is formulated to minimize the total wait time of vessels. The objective function is then solved by simulated annealing and multiple population genetic algorithm (SAMPGA) [7]. The numerical simulation results showed that total wait time reduction from 9788 mins (first come first serve) to 2228 min and 1731 mins for simple genetic algorithm and the proposed algorithm, respectively, for 20 simulated vessels [7]. The limitations of this study are also obvious. Firstly, the study is only considering few number of vessels under simulation and vessel features, such as beam, draft, dwt, etc. In addition, the computational efforts are relatively much larger in the proposed algorithm. On the other hand, unlike the studies of Nishimura et al. and Zhang et al., Dulebenets proposed another way of solving this vessel scheduling problem in shipping operation from both terminal and shipping company perspectives. Since some decisions, which are finally made, are conflicting with each other in nature, the study proposed a combination of multiple conflicting objectives into a single objective function that targets to minimize the total route service costs [8]. This multi-objective mixed integer problem was firstly linearized by discretizing the vessel sailing speed reciprocal, then it was solved by a proposed Global Multi-Objective Optimization algorithm [8]. The study took a shipping line route in Asia-Mediterranean express route as a demonstration. The results showed that the total service cost is significantly reduced on components in both terminals and shipping liners. More in details, Pasha et al. proposed an exact optimization approach for scheduling shipping liner routes between Asia and North America [9]. Most of studies focus on few perspectives, such as purely on shipping liner costs, or terminal service cost, etc. However, Pasha provided a holistic joint-planning model to address this scheduling or planning problem over four tactical-level, which are port service frequency determination, eet deployment, sailing speed optimization and vessel schedule design [9]. According to Pasha's study, a holistic optimization approach was proposed by targeting to maximize total profit obtained from shipping liners within the scopes illustrated as follows: (1) vessel operational cost, (2) vessel chartering cost, (3) port handling cost, (4) port late arrival cost, (5) fuel consumption cost, (6) container inventory cost at sea and ports of call and (7) emission cost at sea and ports of call [9]. The computational experiments were carried out, and the corresponding results showed that BARON could effectively solve the formulated problem within 15 mins, while the maximum optimal gap was within 3%. On the basis of this 20 computational experiments, the problem had already reached 1853 variables and 3063 constraints, the computational efforts had shown to be paid a lot [9]. Therefore, effective computational efforts to solutions may be a new direction to be followed up from this approach.

To the best of our knowledge, majority of optimization scenarios are based on container terminals of ports. Few researches about tanker terminals have been conducted due to the unique characteristics of multiple cargoes, complex operations, etc. Based on our observations, the following factors could be the direct and indirect reasons to the lack of research over tanker terminals:

1. Container terminals deal with standard volumetric containers of cargoes, which could be more effectively operated than liquid cargoes, such as base oil and chemical products. These liquid cargoes are operated by tanker terminals, and they require various and more strict conditions than standard container cargoes. This raises the complexity of optimizing vessels over tanker terminals.
2. Due to the standard volumetric and well pre-planned containers, berth stay of container vessels could be more accurately predictable than that of tanker vessels. This also gives large uncertainty of tanker vessels/terminals for berth stay, and it again makes the optimization of tanker terminals more complicated.
3. Due to the mobility requirements of containers, the containers have to be conveyed into/out of terminal with vehicle fleets. This hinterland scheduling optimization is also taking a significant role of terminal optimization. While tanker terminals are seldom dealing with hinterland mobility. This demonstrates more research opportunities for container terminals than tanker terminals too.

Since plenty of researches are based on container terminal optimization, many performance indicators could be adopted into tanker terminals in ports. Based on the studies of Chung [10], two key performance indicators (KPIs) are included in this study, which are average wait time and average turnaround time.

## Problem Statement

One of many big challenges on scheduling optimization is that tanker vessels/terminals are not coordinative and cooperative enough due to knowledge/information limited and reserved by each individual party or stakeholder. As a result, the scheduling on tanker terminals is not efficient, and it eventually leads to time wastage on anchorage waiting and relevant time wastage. In this study, we proposed a novel intelligent approach to address this scheduling optimization problem by experimenting over consecutive 3-month historical data. In addition, some assumptions are made in this experiment to validate the effectiveness of the proposed approach: 1) certain tanker terminals are dedicated to vessels when visit, 2) certain tanker terminals are inter-exchangeable as buffer terminals to coordinate the vessel visits from heavy-visit terminals to light-visit ones.

## Methodology

In this study, a coordinative scheduling approach is proposed and evaluated. The evaluations are based on two KPIs, such as average wait time and average turnaround time with respect to before and after the proposed approach applied. The experiment is designed by a set of 3-month historical data of tanker vessels/terminals in Singapore port. The objective function covers two weighted KPIs, and an enhanced particle swarm optimization (ePSO) algorithm incorporated into augmented firefly algorithm (AFA) is to minimize the objective function under practical scenarios.

In order to have a solid and comprehensive conclusion on the effectiveness of proposed approach, a number of scenarios and hyper-parameters are introduced and validated. In this study, an overview of analytical flowchart is presented in Figure 4, and the research methodology is demonstrated as follows:

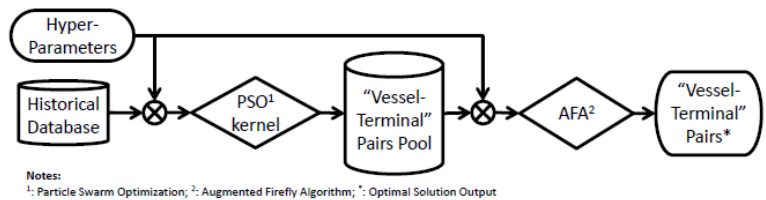

Figure 4: Overview of Analytical Flowchart

a) A reasonable long period of dataset on tanker vessels/terminals is applied in Singapore port. It is from March to May in 2017, and followed by data pre-processing.
b) After analyzing the historical dataset, the statistical of historical average wait time and average turnaround time are benchmarks for effectiveness validation of the proposed approach. The proposed approach is evaluated by two paradigm methods, such as batch method and rolling horizon method.
c) Due to the highly associative relations among particular tanker terminals, certain of less utilized terminals could be treated as buffer terminals for mitigating congestion and reducing time wastage.
d) The hyper-parameter "fixed ratio" is defined as the ratio of fixed operations in terms of timing and terminal over total operations in this study. Another hyper-parameter "randomness ratio" is described as the ratio of operation extent which could be inter-changeable to buffer terminals.
e) Due to practical constraints, fixed and flexible operations among tanker terminals are evaluated by varying hyper-parameters introduced, such as "fixed ratio" and "randomness ratio", to represent different possible practical scenarios.

As the Figure 4 shows above, historical database stores 3-month tanker vessels in Singapore port with data pre-processing. Hyper-parameters are tunable parameters involved in PSO kernel and AFA optimal solution selection. PSO kernel basically samples and generates sub-optimal population of candidates stored as a "vessel-terminal" pairs pool, and AFA further searches for optimal solutions in a bounded high-dimensional searching space. The pseudo-codes of typical PSO [11] enhanced as ePSO and AFA [12] are presented in the following Figure 5 and Figure 6.

```
MaxIter ← Maximum of Iteration
n ← Swarm size
for i = (1: n):
    x_i ← Initial position of i^th particle
    L_{best,i} = x_i
    if h(L_{best,i}) < h(G_{best}):
        G_{best} = L_{best,i}
    v_i ← Initial velocity of i^th particle
endfor

while (t < MaxIter):
    t = t + 1
    Con(x, f, r) ← Initialize fixed and randomness for particles
    for i = (1: n):
        if (x_i satisfies Con(x_i, f, r)):
            v_i ← Update velocity of i^th particle
            x_i(t) = x_i(t − 1) + v_i ← Update position of i^th particle
            if h(L_{best,i}) < h(G_{best}):
                G_{best} = L_{best,i}
endwhile

Output: global best solution, G_{best}
```

Figure 5: Pseudo-code of ePSO

Objective function, $f(\vec{x})$, where $\vec{x} = (x_1, x_2, ..., x_m)$, $\vec{x}_k \in R^m$
Generate initial population $n$ fireflies $\vec{x}_i (i = 1, 2, ..., n)$
Evaluate *Intensity* of fireflies, $I_i = f(\vec{x}_i)$
**while**(not reach stopping criteria)
  **for** $i = 1:1:n$
    **if** $(I_i < I_{max})$
      $\vec{x}_i^{new} = \vec{x}_i^{old} + \alpha \cdot \gamma \cdot (\vec{x}_{max}^{old} - \vec{x}_i^{old}) + \beta \cdot \left[(\Delta B - 1) \cdot s + 1\right] \cdot \varepsilon$
      Update new *Intensity*.
    **else**
      $\vec{x}_i^{new} = \vec{x}_i^{old} + \beta \cdot \left[(\Delta B - 1) \cdot s + 1\right] \cdot \varepsilon$
      Update new *Intensity*.
    **endif**
  **endfor** $i$
**endwhile**
Rank population fireflies and obtain global optimum, $g^*$.
Post-process and visualization.

*Note*:
$\alpha = (0,1]$, Distance coefficient
$\beta = [0,1]$, Randomness coefficient
$\gamma = [0,1]$, Vortex coeffcient
$\varepsilon$ = Uniform / Gaussian distribution
$\Delta B$ = Maximum Boundary Difference
$s = \begin{cases} 0, \text{(Small Region Wandering)} \\ 1, \text{(Large Region Wandering)} \end{cases}$

Figure 6: Pseudo-code of Typical AFA

As the Figure 4 and Figure 5 show, the PSO stage aims to sample and generate sub-optimal "vessel-terminal" pairs pool. The corresponding hyper-parameters of PSO kernel stage are tabulated in Table 1 below:

Table 1: Hyper-parameters of ePSO

| Hyper-parameter | Value | Remark |
|---|---|---|
| $MaxIter$ | 100 | Maximum number of iterations (i.e. number of pairs pool) |
| $n$ | Variable | Number of distinctive vessels |
| $f$ | [0, 0.9] | Fixed ratio of "vessel-terminal" pairs |
| $r$ | [0, 0.9] | Randomness ratio of "vessel-terminal" pairs |

As the Figure 4 and Figure 6 show, the AFA stage targets to deeply search optimal solution in "vessel-terminal" pairs pool based on objective function, $I_i = f(X_i)$, for $i^{th}$ set of pairs in the high-dimensional searching space. For simplicity of tuning different hyper-parameters, the corresponding selected hyper-parameters of AFA stage are tabulated in Table 2 below:

Table 2: Hyper-parameters of AFA

| Hyper-parameter | Value | Remark |
|---|---|---|
| $n$ | Variable | Number of distinctive vessels |
| $\alpha$ | 1 | Distance coefficient |
| $\beta$ | 1 | Randomness coefficient |
| $\gamma$ | 1 | Vortex coefficient |
| $\varepsilon$ | $N \sim (0,1)$ | Gaussian distribution |
| $\Delta B$ | $n$ | Size of pairs pool from ePSO |
| $s$ | 1 | Large region wandering |

## Data Statistics

Given automatic identification systems (AIS) data and port's terminals information, 3-month historical data (March to May of 2017) is prepared. In this section, data statistics are pictured in Table 3 and Figure 7.

Table 3: Data Statistics of Vessels (March - May 2017)

| Item | Value | Remark |
|---|---|---|
| Anchorage | 64 | Number of anchorages |
| Berth | 339 | Number of berths |
| Vessel | 2088 | Number of distinctive vessels |
| Portcall | 4341 | Number of distinctive portcalls |
| Datetime | 01 March 2017 | Starting at 00:00:00 |
| Datetime | 31 May 2017 | Ending at 23:59:59 |

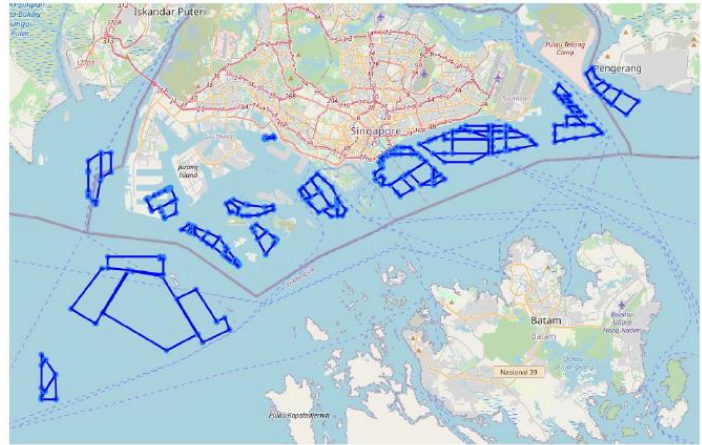

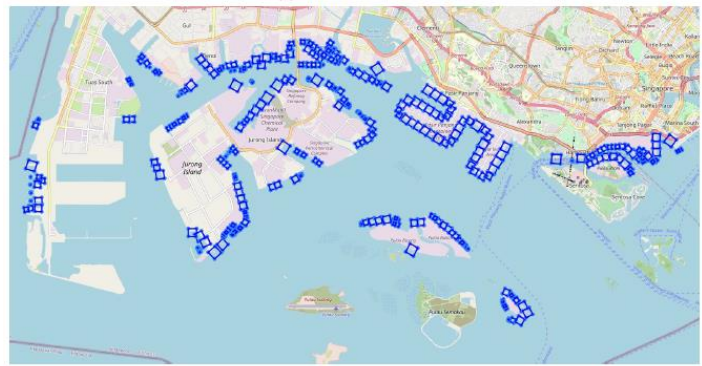

Figure 7: Locations of Anchorages and Berths

## Data Pre-processing

In order to have solid, substantial and comprehensive conclusions, raw data has to be pre-processed before plugging into optimization algorithms. The raw data is stored in a way of large amount of spatial-temporal. The spatial data is geographical longitude and latitude with respect to a designated temporal timestamp. Since the raw data was originated from AIS devices equipped on board, errors could be introduced by apparatus and human being, such as wrong GPS signaling, crews on-board purposely turning off AIS device, etc. Hereby, data pre-processing for standardizing and cleaning the raw data is one of necessary and essential works to complete first.

### Data Standardizing

In this study, the data extracted from database is standardized by information as follows:

a) Location of stay: This covers the longitude and latitude where vessels stay, such as anchorage locations or berth locations of waiting for berthing or conducting operations, respectively.
b) Vessel identity: This includes maritime mobile service identity (MMSI) of vessels for differentiating each particular vessel.
c) Timestamp of stay: This records the starting and ending timestamps of vessels staying at different specific locations.

## Data Cleaning

In this study, the data cleaning is illustrated for the main issues listed as follows:

a) Location drifting: By analyzing the given AIS data, some AIS points are presented as far from their reasonable locations based on their near-term AIS data. This is called drifting in the context of this study as shown in Figure 8. These sudden changes in locations could trigger other errors that propagate in berth stay duration for later processes. Therefore, intelligent algorithms are developed to classify if drifting occurs and output the raw cleaned data thereafter.

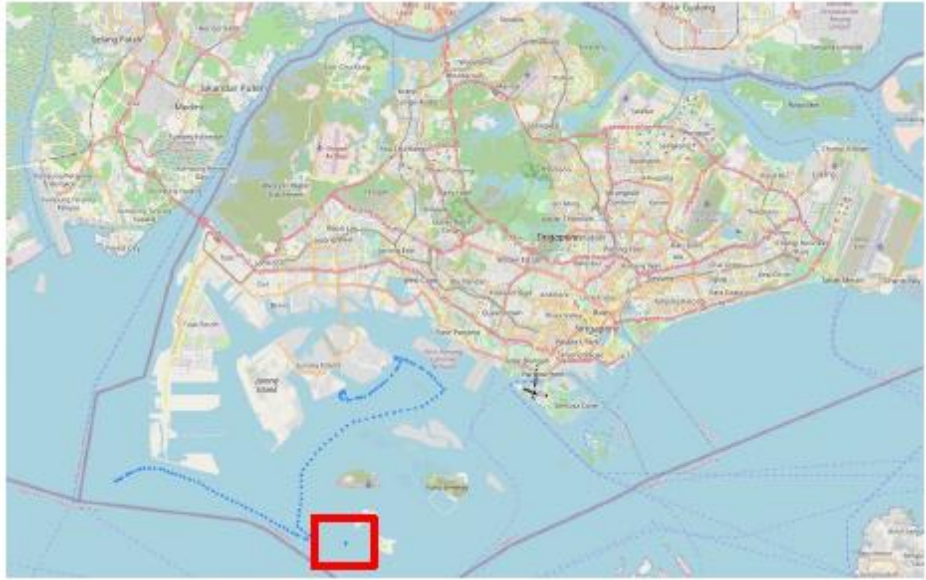

Figure 8: Example Case - Location Drifting

b) Location missing: In particular portcalls of vessels, on-board AIS devices are not always turned on by crews, so some missing gaps could be observed as shown in Figure 9. Based on the missing parts and near-term AIS data, certain missing gaps could be confidently recovered from predictive algorithms, while others have to leave them as they are due to unpredictable behaviors.

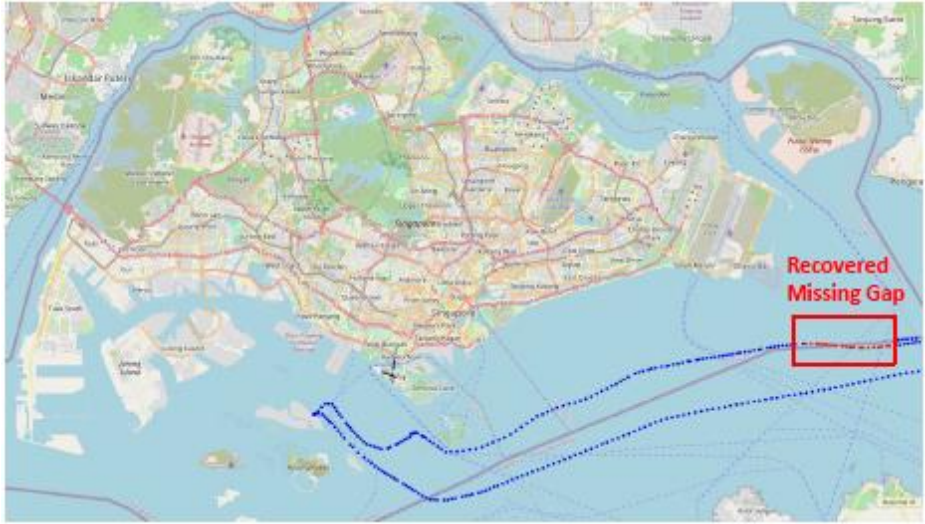

Figure 9: Example Case - Location Missing

## Optimization for Coordinative Scheduling

In this section, the proposed approach with two paradigm methods and scenarios are validated on coordinative scheduling optimization by 3-month historical data. As optimization objectives are to minimize average wait time and turnaround time, the objective function is defined as normalized coefficients $(\alpha, \beta)$ respectively on weighted objectives throughout this study across different paradigms and scenarios. Since the coefficients are normalized and the focus is on turnaround time, minimizing the turnaround time (equivalent to maximizing useful time on operations in port) also reduces the wait time (equivalent to decreasing unused time in port). In this study, for simplicity, $(\alpha, \beta) = (0, 1)$ is selected for this stage of study and the kernel of optimizer is enhanced particle swarm optimization (ePSO) where particle sampling process is leveraged accordingly with two hyper-parameters which are introduced as fixed ratio $(f)$ and randomness ratio $(r)$, given $f, r \in S = \{s | 0 \leq s \leq 0.9, s \in R\}$. By definition, fixed ratio is a parameter that tunes the fixed ratio of "vessel-terminal" pairs. For instance, $f = 0$ means all "vessel-terminal" pairs are not fixed, i.e. a vessel is flexible to re-pair to any other terminal. Similarly, randomness ratio is a parameter that adjusts the randomness ratio of "vessel-terminal" pairs. For example, $r = 0.9$ means that 90% of those flexible "vessel-terminal" pairs are eligible to inter-changeable between current terminals and buffer terminals. In this case, buffer terminals are those terminals that are less utilized and equivalently capable to handle similar cargoes.

### Approach

According to the proposed approach, two paradigm methods, batch method and rolling horizon method, are developed and validated. Their corresponding optimal results are presented in the next section.

a) Batch method: This method aims to optimize 3-month data as a whole batch. The whole 3-month scheduling is treated as one big chunk of job to optimize for optimizers.
b) Rolling horizon method: Instead of taking whole chunk of data at once to feed into the optimizers, this method is to split the 3-month data into weekly rolling horizons. By rolling the horizons of time windows, the optimizers run exponentially faster for each horizon week after week, due to significantly reduced the size of data fed into the optimizers.

### Scenario

With the introduced hyper-parameters $(f, r)$, different types of scenarios could be achieved by manipulating this pair of hyper-parameters. There are many more scenarios investigated, but only some typical scenarios are listed as follows.

a) Given $(f, r) = (0, 0)$: This scenario is to simulate a use-case that is very flexible. All of "vessel-terminal" pairs are flexible, but the buffer terminals are not involved in "vessel-terminal" pairs.
b) Given $(f, r) = (0, 0.9)$: This scenario is to simulate a use-case that is very flexible. All of "vessel-terminal" pairs are flexible, and 90% of those flexible pairs are inter-changeable with buffer terminals.
c) Given $(f, r) = (0.9, 0)$: This scenario is to mimic a use-case that is very rigid. 90% of "vessel-terminal" pairs are fixed and there is no resilience to utilize the buffer terminals.
d) Given $(f, r) = (0.9, 0.9)$: This scenario is to mimic a use-case that is also very rigid. 90% of "vessel-terminal" pairs are fixed, and 90% of those remaining flexible pairs are inter-changeable with buffer terminals.

## Result and Discussion

In this section, results and discussion on different scenarios are presented. Two aspects are studied: (1) optimization effectiveness (i.e. the potential of reduction on time wastage); (2) optimization runtime elapsed. The first part evaluates whether the proposed approach has the capability of addressing this problem effectively, while the second part investigates whether this proposed approach has the advantage capability of near real-time runtime.

### Optimization Effectiveness

Based on the proposed two paradigm methods and scenarios described in the previous section, different combinations of $(f, r)$ scenarios are investigated for average wait time and turnaround time as shown in Figure 10 and Figure 11, respectively.

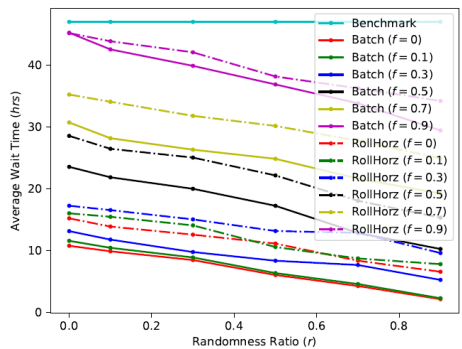
Figure 10: Evaluations of Average Wait Time

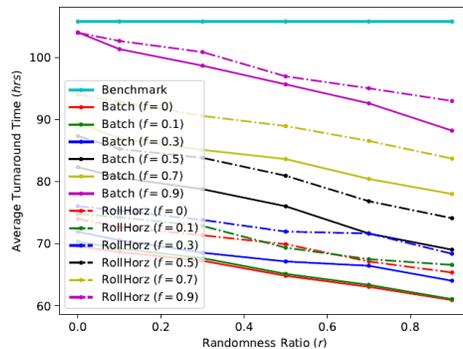
Figure 11: Evaluations of Average Turnaround Time

From Figure 10 and Figure 11, the trends of optimization are homogeneous with discrete $r$ values (from 0 to 0.9) under the discrete $f$ values for both batch and rolling horizon methods. These homogeneous trends decrease average wait time and turnaround time with raising of $r$ values (given an arbitrary $f$ value), while the average wait time and turnaround time decrease with degrading $f$ values (given an arbitrary $r$ value) for both batch and rolling horizon methods.

Comparing between batch and rolling horizon methods, the results show that batch method outperforms rolling horizon method in general. The reason behind this phenomenon is that rolling horizon method is merely to optimize under a given time window which is rolling forward. If the period has rolled forward to the next period, then past schedules are no longer eligible to be further optimized. However, the batch method sees the whole chunk of events as one at each step of scheduling iteration, thus it shows better performance results than rolling horizon method. However, both methods are able to demonstrate that they are capable of reducing average wait time and turnaround time effectively with properly selected hyper-parameters.

According to Figure 10, the maximum reduction of average wait time is about 45 hours and 40.42 hours for batch method and rolling horizon method accordingly. Similarly shown in Figure 11, the maximum reduction of average turnaround time is about 44.87 hours and 40 hours for batch method and rolling horizon method, respectively. These results indicate that there are potentially around 86.0% - 95.5% reduction of average wait time and 38.2% - 42.4% reduction of average turnaround time at the maximum extent. The overall reduction levels across different scenarios are tabulated in Table 4 - 7 for average wait time and turnaround time with respect to batch and rolling horizon methods correspondingly.

Table 4: Reduction of Average Wait Time (Batch)

| $f/r$ | 0 | 0.1 | 0.3 | 0.5 | 0.7 | 0.9 |
|---|---|---|---|---|---|---|
| 0 | 77.1% | 79.0% | 82.0% | 87.2% | 90.9% | 95.5% |
| 0.1 | 75.4% | 77.8% | 81.1% | 86.5% | 90.3% | 95.2% |
| 0.3 | 72.1% | 75.0% | 79.2% | 82.2% | 83.7% | 88.8% |
| 0.5 | 49.9% | 53.5% | 57.5% | 63.3% | 72.7% | 78.2% |
| 0.7 | 34.6% | 40.1% | 44.0% | 47.1% | 54.0% | 59.1% |
| 0.9 | 3.7% | 9.5% | 15.1% | 21.5% | 28.0% | 37.4% |

Table 5: Reduction of Average Turnaround Time (Batch)

| $f/r$ | 0 | 0.1 | 0.3 | 0.5 | 0.7 | 0.9 |
|---|---|---|---|---|---|---|
| 0 | 34.2% | 35.1% | 36.4% | 38.7% | 40.4% | 42.4% |
| 0.1 | 33.5% | 34.6% | 36.0% | 38.4% | 40.1% | 42.3% |
| 0.3 | 32.0% | 33.3% | 35.2% | 36.5% | 37.2% | 39.5% |
| 0.5 | 22.2% | 23.8% | 25.5% | 28.1% | 32.3% | 34.7% |
| 0.7 | 15.4% | 17.8% | 19.5% | 20.9% | 24.0% | 26.3% |
| 0.9 | 1.7% | 4.2% | 6.7% | 9.6% | 12.4% | 16.6% |

Table 6: Reduction of Average Wait Time (Rolling)

| $f/r$ | 0 | 0.1 | 0.3 | 0.5 | 0.7 | 0.9 |
|---|---|---|---|---|---|---|
| 0 | 67.6% | 70.5% | 73.3% | 76.3% | 82.2% | 86.0% |
| 0.1 | 66.0% | 67.1% | 70.1% | 77.5% | 81.5% | 83.4% |
| 0.3 | 63.3% | 64.8% | 68.0% | 72.0% | 72.6% | 79.6% |
| 0.5 | 39.3% | 43.7% | 46.7% | 52.9% | 61.6% | 67.4% |
| 0.7 | 25.0% | 27.5% | 32.4% | 35.8% | 40.9% | 47.0% |
| 0.9 | 3.9% | 6.7% | 10.5% | 18.8% | 22.9% | 27.2% |

Table 7: Reduction of Average Turnaround Time (Rolling)

| $f/r$ | 0 | 0.1 | 0.3 | 0.5 | 0.7 | 0.9 |
|---|---|---|---|---|---|---|
| 0 | 30.0% | 31.3% | 32.5% | 33.9% | 36.5% | 38.2% |
| 0.1 | 29.3% | 29.8% | 31.1% | 34.4% | 36.2% | 37.0% |
| 0.3 | 28.1% | 28.8% | 30.2% | 32.0% | 32.3% | 35.3% |
| 0.5 | 17.4% | 19.4% | 20.8% | 23.5% | 27.4% | 30.0% |
| 0.7 | 11.1% | 12.2% | 14.4% | 15.9% | 18.2% | 20.9% |
| 0.9 | 1.7% | 3.0% | 4.7% | 8.3% | 10.2% | 12.1% |

According to Figure 10, the maximum reduction of average wait time is about 45 hours and 40.42 hours for batch method and rolling horizon method accordingly. Similarly shown in Figure 11, the maximum reduction of average turnaround time is about 44.87 hours and 40 hours for batch method and rolling horizon method, respectively. These results indicate that there are potentially around 86.0% - 95.5% reduction of average wait time and 38.2% - 42.4% reduction of average turnaround time at the maximum extent. The overall reduction levels across different scenarios are tabulated in Table 4 - 7 for average wait time and turnaround time with respect to batch and rolling horizon methods correspondingly.

## Optimization Runtime

Based on the previous experiments on different scenarios, the results of runtime are recorded and compared between batch approach and rolling horizon approach for different corresponding $(f, r)$ scenarios. Since both ePSO and AFA have a polynomial computational complexity, $O(n^k)$, with respect to the problem size, the practical runtime are shown in Figure 12 with respect to completion of optimizing 3-month data throughout all experiments. Basic information of computing device to carry out these experiments is described as follows: CPU (Intel Core i7-6600U@2.60 GHz), RAM (8 GB), OS (Windows 10 Enterprise), HD (Toshiba SSD 256 GB) and platform (Python 3.8).

As Figure 12 shows, the maximum extents of computational efforts with hyper-parameters $(f, r) = (0, 0.9)$ for batch method and rolling horizon method are rounded about 14400 and 1200 seconds, respectively. The rolling horizon method is polynomial faster than batch method through the experiments with the growth of problem size. Based on these results, it is promising to note that balancing between effectiveness and runtime will demonstrate that rolling horizon method has its advantages to cope near real-time optimization with optimal and reasonably optimal solutions obtained.

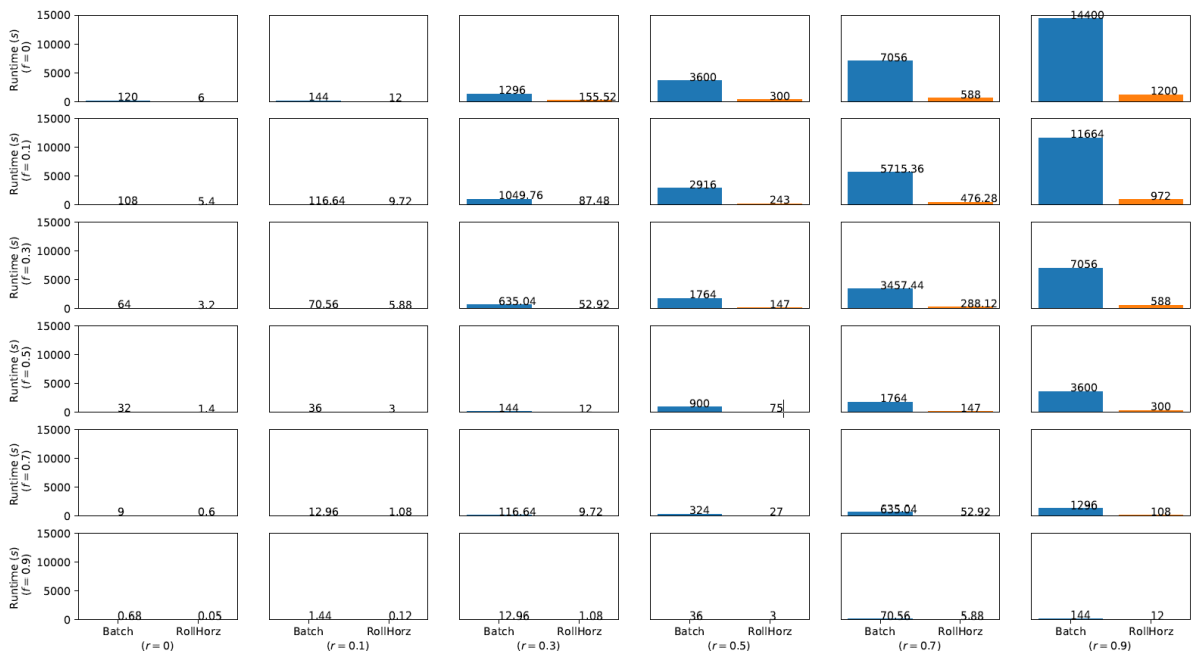

Figure 12: Evaluations of Optimization Runtime

# Conclusion

In this section, conclusion will be drawn, and limitations will be followed up. Lastly, possible future directions will be pointed out.

## Research Conclusion

In this study, the goal is to minimize the average wait time and turnaround time of testing data under a span of 3-month. By addressing this problem, a novel intelligent optimization approach is proposed. The proposed approach composes with enhanced particle swarm optimization (ePSO) as kernel and augmented firefly algorithm (AFA) as high-dimensional search. For optimization paradigms, two methods are proposed, such as batch method and rolling horizon method. Based on the experimental results, rolling horizon method has significant advantages in runtime, compared with batch method.

According to the investigations over different scenarios, the results show that the average wait time could be reduced by 86.0% - 95.5% with respect to historical benchmark, and the average turnaround time could save 38.2% - 42.4% with respect to current benchmark. On the other hand, optimization runtime can be reduced to less than half an hour for rolling horizon method, while 4 hours for batch method over this 3-month designated data at the maximum performance.

This large amount of time saved would directly benefit multiple parties and stakeholders in maritime networks, such as saving fuel/crew/berthing costs for shipping liners, saving time to serve more for ports and terminals, saving chartering costs/delivery time for consignors and consignees, etc. These benefits could eventually let multiple parties and stakeholders gain more profits from these savings. But this economical gain of profits is beyond the scope of this study.

## Research Limitation

This study proposed a novel approach of optimizing vessel scheduling problem and the method is validated through different scenarios. However, many limitations are still involved in this study.

a) Assumption: As mentioned above, introducing hyper-parameters, such as $(f, r)$, there is no distinction between terminals when schedule is optimized by transferring terminals to buffer terminals. However, this is only one assumption for simplicity. In reality, certain terminals are absolutely non-transferable or non-interchangeable.
b) Data: In maritime networks, different types of vessels are voyaging in Singapore water area. However, the data used for this study is merely for tanker vessels. This will limit the overall picture of maritime networks in Singapore port. But the data provided is still successfully capable of showing large potential benefits.

## Future Direction

In future, maritime networks will still play an important role in worldwide economy. High-performance ports will be the common goals to achieve globally. The next step of study would be further cultivating maritime domain, for instance, incorporating other types of vessels that have significantly impacts on port efficiency and congestion. On the other side, optimization algorithms would be further enhanced for both effectiveness and runtime toward different scenarios. The scenarios are also required further investigated for real-kind scenarios, and look forward to real-time optimization with reasonably fast runtime algorithms.

## Acknowledgment

The authors would like to thank Mr. Chua Chye Poh and many others from ShipsFocus Group https://www.shipsfocus.com) for providing helpful discussion and domain knowledge in the field of maritime. The authors also gratefully thank for all kinds of reviews, constructive comments and suggestions.